\documentclass[seceqn,secthm,twocolumn,10pt]{baltzer}
\usepackage{graphicx}

\begin{document}
\sloppy

\begin{frontmatter}  

\title{\Large Exotic Superconducting Phase in CeCu$_2$Si$_2$ Close to Antiferromagnetism : A Cu-NQR Study under Hydrostatic Pressure}

\author[A]{Y.~Kawasaki, K.~Ishida, T.~Mito, C.~Thessieu, G.~-q.~Zheng, Y.~Kitaoka,}      
\author[B]{C.~Geibel and F.~Steglich}
\address[A]{Department of Physical Science, Graduate School of Engineering Science, Osaka University, Toyonaka, Osaka 560-8531, Japan}   
\address[B]{Max-Planck Institute for Chemical Physics of Solids, D-01187 Dresden, Germany}
\runningauthor{}
\runningtitle{}


\begin{abstract}  

We report Cu nuclear-quadrupole-resonance (NQR) results under pressure ($P$) on a homogeneous CeCu$_2$Si$_2$ ($T_{\rm c} = 0.65$ K) that revealed critical magnetic fluctuations at the border to an antiferromagnetic (AF) phase.
This exotic superconducting (SC) phase evolves into the typical heavy-fermion SC phase at minute pressures exceeding $P_{\rm c} \sim$ 0.2 GPa\@.
The nuclear spin-lattice relaxation data in a $P$ range 0.85 -- 2.58 GPa are shown to be accountable by the SCR theory based on the nearly AF Fermi-liquid model.
In $0\leq P< P_{\rm c} \sim 0.2$ GPa, by contrast, we conclude that the exotic SC phase manifests itself under the unconventional normal state where AF waves propagate over a long range without any trace of AF order and thereby the heavy-fermion state breaks up.

\end{abstract}

\end{frontmatter}



There is increasing evidence that a superconducting (SC) phase in cerium-based heavy-fermion (HF) compounds takes place on the border at which an antiferromagnetic (AF) order is suppressed by applying pressure ($P$) to the AF-HF compounds CeCu$_2$Ge$_2$\cite{CeCu2Ge2}, CePd$_2$Si$_2$\cite{CePd2Si2} and CeIn$_3$\cite{CeIn3}.
When a magnetic medium is near the AF phase, AF waves tend to propagate over a long range with low characteristic energy.
Thereby it was argued that the binding of the Cooper pairs could be described in terms of the emission and absorption of fluctuating AF waves (FAFW).
An interplay between the antiferromagnetism and the unconventional superconductivity in the cerium-based HF superconductors may share some common aspects with that in other strongly-correlated-electron systems including even high-$T_{\rm c}$ cuprates and organic systems.
In the former, a crossover from an AF insulator to an unconventional SC state is caused by doping carriers, while in the latter it occurs by either a slight chemical substitution or an application of the pressure.
The presence of FAFW in the metallic state is believed as one of key features for the occurrence of superconductivity in these strongly-correlated-electron systems.
Possible mechanism for the superconductivity that should be different from the conventional electron-phonon mediated one is currently important issue to resolve over a decade.

The underlying HF-superconductor CeCu$_2$Si$_2$ is one of the most suitable compounds to address the above issue, since it lies at the vicinity of an AF phase at ambient pressure ($P = 0$) \cite{gegenwart,trovarelli,ishida99}.
The observation of magnetic {\it phase A} at magnetic fields exceeding an upper critical field $H_{\rm c2}$ is proposed to be of a spin-density-wave (SDW)-type \cite{bruls,gegenwart}.
This evidences the closeness to the AF phase \cite{ishida99,Kitaoka}.
$T_{\rm A}$ that marks the transition into {\it phase A} at $P$ = 0 is close to the SC transition temperature $T_{\rm c}$ in homogeneous poly- and single-crystal samples.
It is, however, under a debate whether {\it phase A} at zero field ($B = 0$) is of a static long-range order \cite{gegenwart,trovarelli} or associated with FAFW that have extremely low characteristic frequencies comparable to an NQR frequency ($\sim$ 3 MHz) \cite{ishida99,Kitaoka,feyerherm}. 
{\it Phase A} was denoted as {\it critical magnetic phase} in the previous paper \cite{ishida99}, since the NQR experiments did not evidence any trace for the onset of AF order down to 0.012 K\@.
Recently, the Cu-NQR experiments on CeCu$_2$(Si$_{1-x}$Ge$_x$)$_2$ have shown that FAFW evolve into an AF phase, once slight Ge substitution for Si expands its unit cell volume.
{\it Phase A} that was suggested to exist even at $B = 0$ \cite{trovarelli} was shown to be neither an AF- nor SDW-type phase \cite{Kitaoka}.

Furthermore, thorough investigations on a series of near stoichiometric samples
succeeded to clarify an intimate crossover from the AF to SC phase as follows.
(1) In type-I Ce$_{0.99}$Cu$_{2.02}$Si$_2$ (Ce0.99) where $T_{\rm c} \sim$ 0.65 K $ < T_{\rm A} \sim 0.75$ K, the Cu-NQR measurements revealed that FAFW develop rapidly upon cooling towards $T_{\rm A}$, but any static magnetic ordering was not identified down to 0.012 K \cite{ishida99}.
A clear indication for the onset of SC transition was evidenced from a rapid decrease in the nuclear spin-lattice relaxation rate $1/T_1$ just below $T_{\rm c}$\@.
Peak at $T_{\rm c}$ and large enhancement in $1/T_1T$ well below $T_{\rm c}$, on the other hand, pointed to the presence of FAFW even in the SC state. These results ensure that Ce0.99 is just on the border to the AF phase.
(2) In type-II CeCu$_{2.05}$Si$_2$ (Ce1.00) where $T_{\rm A} \sim T_{\rm c}$\@.
The onset of a typical HF-SC transition expels FAFW below $T_{\rm c}$ \cite{bruls,ishida99,steglich}.
The observation of $1/T_1 \propto T^3$-like dependence also evidences the typical behavior for HF superconductors as a consequence of the suppression of FAFW \cite{ishida99}.
In type-III Ce$_{1.025}$Cu$_{2.02}$Si$_{2}$ (Ce1.025) where $T_{\rm c} \sim 0.65$ K $ > T_{\rm A}$ \cite{ishida99}, typical HF-SC signatures were observed without any clear sign for the development of FAFW\@.

We remark that FAFW were also reported in high-$T_{\rm c}$ cuprates La$_{2-x}$Sr$_x$CuO$_4$ through the {\it wipe-out} phenomenon of Cu-NQR intensity in Nd- and Eu-doped systems and the underdoped compounds in a concentration range $1/16 < x < 1/8$ which exhibit spin and charge stripe orders or fluctuations \cite{Hunt,Curro}.
Characteristic frequencies of these fluctuations around 10 K are estimated to be in a range 5--50 MHz, comparable to those of FAFW in the type-I Ce0.99\@.
FAFW with very low frequencies thus exist at the crossover region from the AF to SC phase in strongly-correlated-electron systems.
Thereby a detailed study to characterize FAFW is quite important subject to elucidate the mechanism of unconventional superconductivity and an interplay between antiferromagnetism and superconductivity. 

In order to gain deep insight into the exotic but complicated magnetic and SC properties in the type-I Ce0.99, we report extensive Cu-NQR studies under hydrostatic pressure.
Temperature ($T$) dependence of NQR intensity and nuclear spin-lattice relaxation rate $1/T_1$ were measured in a $P$ range 0 -- 2.58 GPa and a $T$ range 0.1 -- 50 K\@.
Remarkable results of this study are (1) FAFW are strongly suppressed at minute pressures exceeding $P_{\rm c} \sim$ 0.2 GPa\@.
(2) Correspondingly, unconventional $1/T_1$ behaviors that reveal a peak at $T_{\rm c}$ and a large enhancement in $1/T_1T$ well below $T_{\rm c}$, evolve into a typical HF-SC one that exhibits $1/T_1 \propto T^3$-like dependence as in the type-II and -III\@.
(3) Reasonable fit of the SCR theory to the $1/T_1T(P)$ data in a $P$ range 0.85 -- 2.58 GPa enables us to deduce several parameters characterizing the development of AF spin fluctuations (AFSF).
We suggest a {\it breakup} of HF state because of the persistence of FAFW in $P < P_{\rm c}\sim 0.2$ GPa where strongly interacting low-energy degrees of freedom become dominant.

Detailed-preparation method of the homogeneous polycrystal Ce0.99 was reported in the literature \cite{modler}.
We used the same sample as in the previous work \cite{gegenwart,ishida99,modler,feyerherm}.
The EPMA (electron probe microanalysis) result showed that a main phase exhibits an ideal Ce:Cu:Si stoichiometry of 1:2:2, where the inhomogeneity in composition is less than $1\%$ \cite{feyerherm}.
Diamagnetic signal observed upon cooling below $T_{\rm c} \sim 0.65$ K was recently reported to reach a value for an ideal Meissner shielding at a lowest temperature \cite{gegenwart2}.
High-frequency ac-susceptibility (ac-$\chi$) data were also measured by using an in-situ NQR coil.
The bulk SC nature in Ce0.99 is corroborated by a comparable size at low temperatures among the ac-$\chi$ data at various values of pressure that are presented in the inset of Fig.~1\@.
The sample were moderately crushed into grains with diameters larger than 100 $\mu$m in order to avoid some crystal distortion.
The previous study showed that the full-width at half-maximum (FWHM) of the Cu-NQR spectrum in Ce0.99 is as narrow as $\sim$ 14 kHz that is comparable to those in the type-II poly- and single-crystal samples, confirming {\it its high quality in a microscopic scale} \cite{ishida99}.
Hydrostatic pressure was applied by utilizing a CuBe piston-cylinder cell with an equal volume of fluorinert (type 77 and 70) as a pressure-transmitting medium.
To calibrate a pressure value at a sample position at low temperatures, $P$-induced $T_{\rm c}$ variation of Sn powder was measured by high-frequency ac-$\chi$ measurement using the in-situ NQR coil.
The Cu-NQR spectrum was obtained by the Fourier transformed technique of spin-echo signal.
$1/T_1$ was measured by the conventional saturation-recovery method in a $T$ range 0.1 -- 50 K\@.
The $P$-induced variation of the $^{63}$Cu-NQR spectrum at 4.2 K indicates that the NQR frequency $\nu_{\rm Q}$ increases linearly.
This result suggests that an increase in the lattice density gives rise to a linear increase of the electric field gradient at the Cu site in proportion to $\nu_{\rm Q}$.
An increasing rate of $\nu_{\rm Q}$, $d\nu_{\rm Q}/dp = 9.52$ Hz/bar is in good agreement with the value in the previous report \cite{kitaoka95}.
The gradual increase of FWHM with increasing $P$ assures that a possible $P$ distribution is about $\pm$ 0.03 GPa at $P = 0.29$ GPa and $\pm$ 0.06 GPa at $P$ = 2.58 GPa\@.

\begin{figure}
\begin{center}
\includegraphics[width=7.5cm]{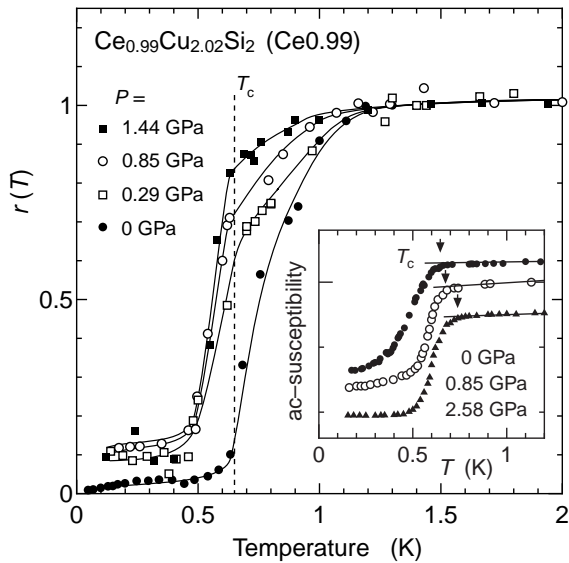}
\caption[]{$T$ dependence of the spin-echo intensity $I$ multiplied by $T$, $r(T) = (I \times T)/[I(4.2) \times 4.2]$ normalized by the value at 4.2 K at various values of pressure.
The inset shows the in-situ ac-susceptibility results at $P=$ 0, 0.85 and 2.58 GPa\@.}
\end{center}
\end{figure}

Fig.~1 displays the $T$ dependence of the NQR intensity ($I$) of Ce0.99 multiplied by temperature, $r(T) = (I\times T)/[I(4.2)\times 4.2]$ normalized by the value at 4.2 K\@.
A $T_{\rm c}$ that was determined by the onset of SC transition through the in-situ ac-$\chi$ measurement stays nearly a constant value of $T_{\rm c} \sim 0.65$ K up to $P$ = 1.44 GPa (see the inset in Fig.~1).
This $T_{\rm c}$ value is indicated by dotted line in the main figure.
Note that $T_{\rm c}$ starts to increase up to 0.75 K at $P$ = 2.58 GPa\@.
It was previously reported that the $r(T)$ at $P = 0$ decreases markedly from $r(T) = 1$ above $T_{\rm m} \sim 1.2$ K down to $r(T_{\rm c}$ = 0.65 K) = 0.1 \cite{ishida99}.
This arises because FAFW develop at very low frequencies that are comparable to NQR frequencies.
In this case, $T_1$ and/or $T_2$ (spin-echo decay time) are estimated as $T_1 \sim T_2 \sim$ 0.14 $\mu$sec from the $\mu$SR results \cite{feyerherm}, which is too short to observe the NQR signal \cite{ishida99}.
Note, on the other hand, a sharp reduction in $r(T)$ below $T_{\rm c}$ at $P$ = 1.44 GPa is due to the SC diamagnetic shielding effect of rf field for the NQR experiment.
It should be noted that $T_{\rm A} \sim 0.75$ K deduced from the $C(T)$ measurements coincides with a temperature where the decreasing rate $dr(T)/dT$ exhibits a maximum.

The $r(T_{\rm c}) = 0.1$ at $P$ = 0 increases up to 0.85 at $P$ = 1.44 GPa (also see Fig.~4a).
Furthermore, increasing $P$ makes $dr(T)/dT$ smaller.
These results indicate that applying the pressure suppresses FAFW, reducing $T_{\rm A}$\@.
A sudden evolution from the type-I to type-II phase ($T_{\rm A}\sim T_{\rm c}$) occurs in a narrow $P$ range 0 -- 0.3 GPa\@.
This event leads to an emergence of typical HF-SC state in 0.3 GPa $< P$\@.
The $r(T)$ at $P = 2.58$ GPa is not reduced at all down to $T_{\rm c}$, followed by a marked decrease below $T_{\rm c}$\@.
This result assures that the steep reduction in $r(T)$ between $T_{\rm c} \sim$ 0.65 K and $T_{\rm m}$ = 1.2 K at pressures lower than $P = 2.58$ GPa is due to the development of FAFW at very low frequencies.

\begin{figure}
\begin{center}
\includegraphics[width=7.5cm]{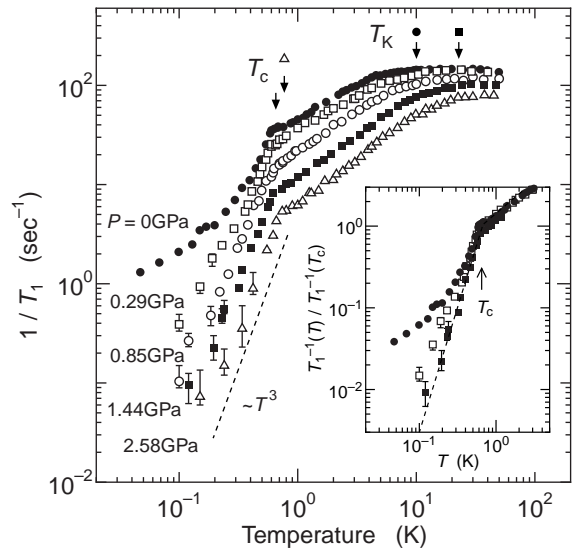}
\caption[]{Nuclear spin-lattice relaxation rate, $1/T_1$ versus $T$ plots at various values of pressure.
The inset shows the $T$ dependence of $1/T_1$ normalized by the value at $T_{\rm c}$, $T_1^{-1}(T)/T_1^{-1}(T_{\rm c})$ along with the data in Ce1.025 at $P=0$ that are presented by dash line.}
\end{center}
\end{figure}

Figure 2 displays the $P$-induced variation in $1/T_1$ in a $P$ range 0 -- 2.58 GPa and a $T$ range 0.1 -- 50 K where $1/T_1$ is determined by a single component.
In the entire $T$ range, $1/T_1$ is suppressed with increasing $P$\@.
The $1/T_1$ results in the low $T$ region reflect the $P$-induced evolution of the ground state.
At $P$ = 0, $1/T_1$ shows a small hump at $T_{\rm c}$\@, followed by a rapid decrease below $T_{\rm c}$ due to the onset of SC transition.
By contrast, its $T$ decreasing behavior becomes much more moderate than the $1/T_1\propto T^3$-like behavior in Ce1.00 and Ce1.025\@.
In the inset of Fig.~2, $1/T_1$'s at $P$ = 0, 0.29 and 1.44 GPa normalized by the value at $T_{\rm c}$, $T_1^{-1}(T)/T_1^{-1}(T_{\rm c})$ are plotted against $T$ along with the data in Ce1.025 at $P$ = 0 that is presented by dash line.
It is clear that an application of $P$ = 0.29 GPa is enough to suppress the large enhancement of $1/T_1$ at $P$ = 0, although $T_{\rm c}$ remains almost unchanged.
We conclude that the exotic $T_1$ behaviors observed at $P = 0$ are relevant to the existence of FAFW that are suppressed at minute pressures exceeding $P =$ 0.29 GPa\@.

We next deal with the pressure effect in the normal state.
Note that a value of $T$-independent $1/T_1$ in the high-$T$ region decreases and a temperature $T^*$, below which $1/T_1$ starts to decrease from the constant $1/T_1$, shifts to a higher $T$ as $P$ increases.
These results indicate that an exchange interaction between $4f$- and conduction-electron spins, $J_{cf}$, increases with $P$, since the value of $1/T_1$ = const is related to $1/J_{cf}^2$ and $T^*$ is empirically related to a Kondo temperature defined by $T_{\rm K} \propto \exp(-W/J_{cf})$.
Here $W$ is a width of conduction band.
The $P$ dependence of $ T^*\sim T_{\rm K}$ is shown in Fig.~4c\@.

A nearly AF Fermi liquid (NAFL) model has been proposed for the understanding of unusual normal and SC properties near the AF phase in HF compounds \cite{nakamura}.
It is interesting to examine if $1/T_1T(P)$ data are accountable by the SCR theory based on the NAFL model \cite{moriya}.
Parameters in the theory is anticipated to be addressed by the experiment.
A spectral width in the energy space of AFSF spectrum $T_0$ is assumed to scale to the effective HF bandwidth $T_{\rm K}$\@.
Figure 3 shows the $t$ dependence of $1/T_1t$ at $P$ = 0, 0.85 1.14, 1.44 and 2.58 GPa, where $t = T/T_{\rm K}$\@.

In the SCR scheme, $1/T_1t$ is calculated based on the following formula \cite{ishigaki};
\[
\frac{1}{T_1t} = \frac{\hbar\gamma_{\rm n}^2A_{\rm hf}^2}{2\pi T_{\rm a}}\frac{3\pi}{4\sqrt{y(y_0,y_1,t)}}
\]
where $\gamma_{\rm n}$ is the gyromagnetic ratio of $^{63}$Cu nucleus and $A_{\rm hf}$ is the hyperfine-coupling constant, $\sim 1.6$ kOe/$\mu_{\rm B}$ in this compound \cite{ueda}.
$T_{\rm a}$ is the characteristic width in the momentum space of AFSF spectrum (in an energy scale) and $y_0$ is proportional to an inverse staggered susceptibility $1/\chi_Q$ at $T = 0$, then $y_0 = 0$ at an AF quantum critical (QC) point.
$y_1$ is the parameter to govern the mode-mode coupling of AFSF\@.
Here $y_1 = 4$ is fixed according to the $C(T)$ results under the pressure \cite{gegenwart}.
The reduced inverse susceptibility $y$ is calculated from
\[
y=y_0+\frac{3}{2}y_1\int_0^{x_{\rm c}}dxx^2\left[\ln{u}-\frac{1}{2u}-\psi(u)\right]
\]
where $u = (y+x^2)/t$, $\psi(u)$ is the digamma function and $x_{\rm c} \sim 1$ is the reduced cut-off wave vector \cite{nakamura}.

\begin{figure}
\begin{center}
\includegraphics[width=7.5cm]{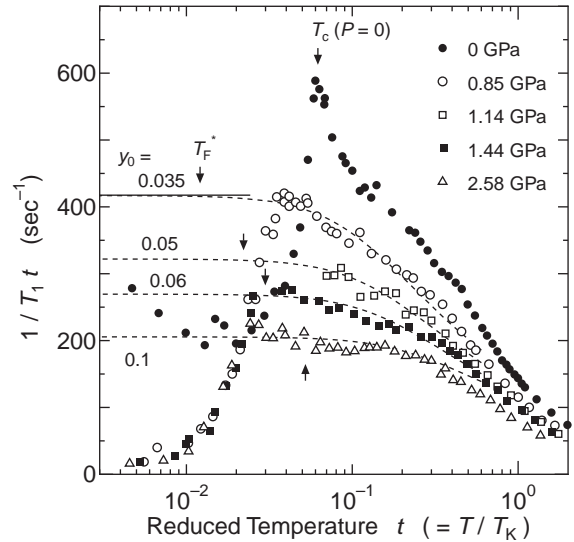}
\caption[]{$1/T_1t$ versus $t = T/T_{\rm K}$ plots at various values of pressure.
Here $t = T/T_{\rm K}$ the reduced temperature divided by the effective HF bandwidth $T_{\rm K}$\@.
Dash lines are fits of the SCR theory to $1/T_1t$ at various values of the parameter $y_0$ and a fixed $y_1$ = 4.
Note that $y_0 \propto \chi_Q^{-1}(T = 0)$ where $\chi_Q(T = 0)$ is the staggered susceptibility and $y_1$ the parameter to govern the mode-mode coupling of AF spin fluctuations.}
\end{center}
\end{figure}

Theoretical curves that are drawn by dash lines in Fig.~3 are consistent with the experimental data in $P >$ 0.85 GPa and $T_{\rm c} < T < 0.5 T^*$\@.
A $T_1t$ = constant behavior in these curves in a low-$T$ range allows us to tentatively deduce another parameter, an effective Fermi temperature $T_{\rm F}^*$ as marked in the figure.
Figs.~4b and 4c show the $P$-induced variations of $y_0$, $T_{\rm K}$ and $T_{\rm F}^*$ along with $T_{\rm c}$, respectively.
The value of $y_0 = 0.1$ at $P$ = 2.58 GPa decreases monotonically to $y_0 \sim 0.035$ at $P$ = 0.85 GPa\@.
Interestingly, a tentative linear fit to five $y_0$ data is extrapolated to $y_0\rightarrow 0$ at $P_{\rm c}(y_0) \sim$ 0 -- 0.2 GPa\@.
This is consistent with the $C(T)$ and $\rho(T)$ results that found a critical pressure value $P^* \sim$ 0.1 GPa exceeding which {\it phase A} is no longer traced by macroscopic measurements \cite{gegenwart}.
$T_{\rm F}^*$ decreases significantly from $\sim$ 1.7 K at $P = 2.58$ GPa to $\sim$ 0.2 K at $P$ = 0.85 GPa\@.
Interestingly, $T_{\rm F}^*$ crosses $T_{\rm c} = 0.65$ K at $P \sim$ 1.5 GPa and a tentative linear fit to five $T_{\rm F}^*$ data is extrapolated to $T_{\rm F}^* \rightarrow 0$ at $P_{\rm c}(T_{\rm F}^*) \sim$ 0.6 GPa\@.
This is also consistent with the fact that the three dimensional AF-QC behaviors emerge in the $C(T)$ and $\rho(T)$ data in the normal state in $P \leq 0.6$ GPa \cite{gegenwart}.

\begin{figure}
\begin{center}
\includegraphics[width=5.5cm]{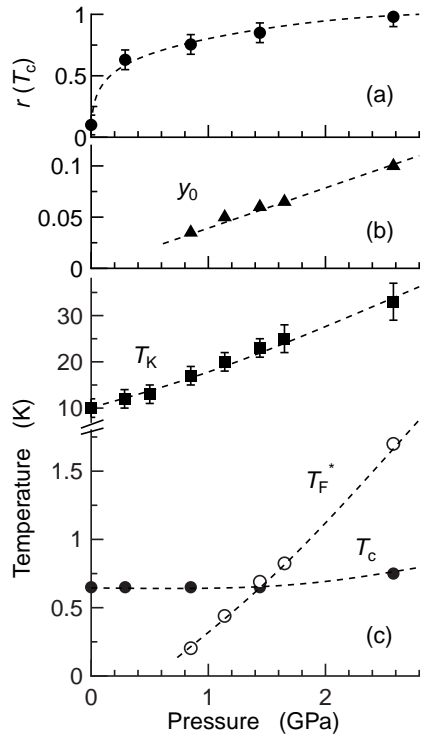}
\caption[]{$P$ dependence of various quantities, (a) $r(T_{\rm c}) = [I(T_{\rm c}) \times T_{\rm c}]/[I(4.2)\times 4.2$], (b) $y_0 \propto \chi_Q^{-1}(T = 0)$, (c) $T_{\rm K}$ (closed boxes), $T_{\rm c}$ (closed circles), and an effective Fermi temperature, $T_{\rm F}^*$ (open circles).}
\end{center}
\end{figure}

The present results are summarized in Figs.~4\@.
The $r(T_{\rm c})$ increases rapidly in $0 < P < P_{\rm c}(y_0) \sim $ 0.2 GPa\@.
In addition, the fact that the $r(T)$ at $P = 2.58$ GPa is not decreased at all down to $T_{\rm c}$ reveals the dramatic suppression of FAFW with the pressure.
Eventually, the exotic SC behaviors are replaced by the typical HF-SC one's that were observed in Ce1.00 and Ce1.025\@.
This evolution in the SC characteristics is accompanied with the increases in $y_0$ and the HF bandwidth $T_{\rm K}$ as $P$ increases as seen in Figs.~4b and 4c.
The normal-state magnetic properties in 0.85 GPa $< P$ and $T_{\rm c} < T < 0.5T^*$ are well described in terms of the SCR theory based on the NAFL model.
It is surprising that $T_{\rm c}$ stays almost constant, nevertheless the AFSF parameters are significantly changed in $0 < P < 1.44$ GPa as seen in Fig.~4c\@.
In the homogeneous type-I Ce0.99 in $0 < P < P_{\rm c}(y_0) \sim $ 0.2 GPa, the exotic SC state manifests itself under the unconventional normal state where FAFW propagates over a long range and thereby the HF state breaks up.

In conclusion, we have found that the exotic SC state in the homogeneous CeCu$_2$Si$_2$ that is at the border to the AF phase evolves into the typical HF-SC state at minute pressures exceeding $P_{\rm c} \sim$ 0.29 GPa\@.
At pressures higher than $P = 0.85$ GPa, the magnetic anomalies have been shown to be accountable based on the nearly AF Fermi-liquid model.
We suggest, on the other hand, at pressures smaller than $P_{\rm c} \sim$ 0.2 GPa that the exotic SC state arises because of the existence of critical magnetic fluctuations or slowly FAFW rather than in spite of them.
In this context, we propose that the magnetic-field induced {\it phase A} is hindered behind this exotic SC state that coexists with FAFW\@.
Possible interpretation of these unconventional SC and magnetic phases in the type-I and - II would involve more exotic but conceptual thinking rather than a conventional picture based on the AF Fermi-liquid model incorporating only low-frequency modes of AFSF\@.

This work was supported by the COE Research (10CE2004) in Grant-in-Aid for Scientific Research from the Ministry of Education, Sport, Science and Culture of Japan.


%
\end{document}